\documentclass[11pt]{article}
\begin{document}

\noindent
\Large
{\bf BOHMIAN PARTICLE TRAJECTORIES IN RELATIVISTIC FERMIONIC
QUANTUM FIELD THEORY}
\normalsize
\vspace*{1cm}

\noindent
{\bf Hrvoje Nikoli\'c}

\vspace*{0.5cm}
\noindent
{\it
Theoretical Physics Division \\
Rudjer Bo\v{s}kovi\'{c} Institute \\
P.O.B. 180, HR-10002 Zagreb, Croatia \\
E-mail: hrvoje@thphys.irb.hr}

\vspace*{2cm}

\noindent
The de Broglie--Bohm interpretation of 
quantum mechanics and quantum field theory 
is generalized in such a way that it
describes trajectories of relativistic fermionic 
particles and antiparticles 
and provides a causal description of the processes
of their creation and destruction. 
A general method of causal interpretation of quantum systems 
is developed and applied to a causal interpretation 
of fermionic quantum field theory 
represented by c-number valued wave functionals. 
\vspace*{0.5cm}

\noindent
Key words: de Broglie--Bohm interpretation, particle trajectory,
fermion, quantum field theory.

\section{INTRODUCTION}

The de Broglie--Bohm (dBB) interpretation of
quantum mechanics (QM) and
quantum field theory (QFT)
\cite{bohm,bohmPR1,bohmPR2,holPR,holbook} was
invented with the intention to provide 
a causal interpretation of the conventional 
QM and QFT. Yet, the current form of the 
dBB interpretation of relativistic fermions still 
does not achieve this goal. First, the Dirac 
spinor is interpreted as a guiding wave only for  
one kind of corpuscles, making no difference 
between particle and antiparticle trajectories 
\cite{hol92,dew,holbook}. Second, QFT 
is not a part of the current form 
of the dBB interpretation of fermions. There exists 
a causal interpretation of fermionic fields in terms of 
quantum rotators \cite{hol_pla,holbook}, but this 
interpretation works only for free fermionic fields.
Besides, the causal interpretation of fermionic fields 
and the interpretation in terms of causal 
trajectories of corpuscles 
are viewed as two mutually incompatible interpretations.
A recent approach \cite{col}, based on earlier work \cite{bell1,bell2}, 
partially resolves these problems, but still makes no difference
between particle and antiparticle trajectories.

A solution of similar problems for bosons have recently been proposed 
in Ref.~\cite{nikoldbb}. In that paper, 
particle trajectories are determined 
by using the concept of particle current \cite{nikol1,nikol2,nikol3}
and multiparticle wave functions that result from QFT. 
A causal interpretation of particle creation and destruction  
is given in terms of causally evolving fields and 
new hidden variables - particle effectivities. 
Both the particle positions and the fields are considered as 
beables (i.e., elements of reality). The 
statistical predictions of the theory are compatible with those
of the standard bosonic QFT. 
However, despite the compatibility, the predictions of the theory 
are {\em not equivalent} to those of the standard bosonic QFT.
Instead, the theory has predictions on the statistical distribution 
of particle positions, on which, in general, the standard bosonic QFT
does not have clear predictions. 

In this paper we generalize the results of Ref.~\cite{nikoldbb}
to fermionic fields. 
It should be considered mainly as a technical generalization.
Consequently, the interpretational 
issues are not discussed in detail as they are already 
extensively discussed in Ref.~\cite{nikoldbb}.
In Sec.~\ref{S2} we give a causal interpretation
of the Dirac equation in terms of particle and antiparticle 
trajectories. In Sec.~\ref{S3} we use the conventional 
interacting fermionic QFT in the Schr\"odinger picture 
\cite{flor,kief,hall} to 
construct multiparticle wave functions and the equations for the 
corresponding particle and antiparticle trajectories. 
In Sec.~\ref{S4}, to each fermionic state 
represented by a Grassmann valued wave functional 
we attribute a corresponding c-number valued wave functional.
In Sec.~\ref{S5} we develop a general method of causal 
interpretation of an arbitrary quantum system in which a 
c-number valued wave function satisfies a Schr\"odinger equation.
In Sec.~\ref{S6} we use the results of Secs.~\ref{S4} 
and \ref{S5} to construct a 
causal interpretation of fermionic QFT and a causal interpretation of
particle creation and destruction, 
in a way similar to that for bosonic fields.
The conclusions are drawn in Sec.~\ref{S7}.

\section{CAUSAL INTERPRETATION OF THE DIRAC \\ EQUATION}
\label{S2}

The Dirac equation in the Minkowski metric 
$\eta_{\mu\nu}\!=\!{\rm diag}(1,-1,-1,-1)$ is
\begin{equation}\label{f1}
(i\gamma^{\mu}\partial_{\mu}-m)\psi(x)=0,
\end{equation}
where $x=(x^0,x^1,x^2,x^3)=(t,\bf{x})$.
The general solution of ({\ref{f1}) 
can be written as 
\begin{equation}\label{f2}
\psi(x)=\psi^{(P)}(x)+\psi^{(A)}(x),
\end{equation}
where the particle and antiparticle parts can be expanded as 
\begin{equation}\label{f3}
\psi^{(P)}(x)=\sum_k b_k u_k(x), \;\;\;
\psi^{(A)}(x)=\sum_k d_k^* v_k(x), 
\end{equation}
respectively. Here $u_k$ ($v_k$) are positive (negative) frequency 
4-spinors that, together, form a complete orthonormal set of 
solutions to (\ref{f1}). The label $k$ is an abbreviation for the 
set $({\bf k},s)$, where ${\bf k}$ is a 3-vector related to a 
Fourier expansion and $s=\pm 1/2$ is the spin label.

The interpretation of negative frequency spinors as antiparticles 
(i.e., positrons) above is usually considered as a part of second 
quantization, i.e., QFT. Indeed, QFT is discussed in the next section.
However, by using the St\"uckelberg-Feynman theory, 
such an interpretation 
is meaningful even without explicit introduction of QFT \cite{bjdr}.  

By introducing the quantities 
\begin{eqnarray}\label{f4}
& \Omega^{(P)}(x,x')=\displaystyle\sum_k u_k(x) u_k^{\dagger}(x'), & 
\nonumber \\
& \Omega^{(A)}(x,x')=\displaystyle\sum_k v_k(x) v_k^{\dagger}(x'), &
\end{eqnarray}
we can extract $\psi^{(P)}$ and $\psi^{(A)}$ from $\psi$ 
using
\begin{eqnarray}\label{f5}
& \psi^{(P)}(x)=\displaystyle\int d^3x' \,\Omega^{(P)}(x,x')\psi(x'), &
\nonumber \\
& \psi^{(A)}(x)=\displaystyle\int d^3x' \,\Omega^{(A)}(x,x')\psi(x'), &
\end{eqnarray}
where $t=t'$. We introduce the particle and antiparticle 
currents defined as
\begin{equation}\label{f6}
j_{\mu}^{(P)}=\bar{\psi}^{(P)} \gamma_{\mu}\psi^{(P)}, \;\;\;
j_{\mu}^{(A)}=\bar{\psi}^{(A)} \gamma_{\mu}\psi^{(A)},
\end{equation}
respectively,
where $\bar{\psi}=\psi^{\dagger}\gamma_0$. Since $\psi^{(P)}$ 
and $\psi^{(A)}$ separately satisfy the Dirac equation (\ref{f1}), 
the currents $j_{\mu}^{(P)}$ and $j_{\mu}^{(A)}$ are separately 
conserved: $\partial^{\mu}j_{\mu}^{(P)}=\partial^{\mu}j_{\mu}^{(A)}=0$.
Therefore, we postulate that trajectories of particles and antiparticles 
are given by
\begin{equation}\label{f7}
\displaystyle\frac{d{\bf x}^{(P)}}{dt}=
\frac{{\bf j}^{(P)}(t,{\bf x}^{(P)})}
{j_0^{(P)}(t,{\bf x}^{(P)})}, \;\;\;
\displaystyle\frac{d{\bf x}^{(A)}}{dt}=
\frac{{\bf j}^{(A)}(t,{\bf x}^{(A)})}      
{j_0^{(A)}(t,{\bf x}^{(A)})}, 
\end{equation}
respectively, where ${\bf j}=(j^1,j^2,j^3)$.

\section{PARTICLE AND ANTIPARTICLE TRAJECTORIES IN \\ FERMIONIC QFT}
\label{S3}

In fermionic QFT, the coefficients $b_k$ and $d^*_k$ in (\ref{f3}) become 
anticommuting operators. The operators $\hat{b}_k^{\dagger}$ 
($\hat{d}_k^{\dagger}$) create particles (antiparticles), while 
$\hat{b}_k$ and $\hat{d}_k$ annihilate them. In the Schr\"odinger 
picture, the field operators $\hat{\psi}({\bf x})$ and 
$\hat{\psi}^{\dagger}({\bf x})$ satisfy the anticommutation relations
\begin{equation}\label{f8}
\{ \hat{\psi}_a({\bf x}), \hat{\psi}_{a'}^{\dagger}({\bf x}') \}
= \delta_{aa'}\delta^3({\bf x}-{\bf x}'), 
\end{equation}
while other anticommutators vanish. (Here $a$ is a spinor index.)
These anticommutation relations can be represented by 
\begin{eqnarray}\label{f9}
& \hat{\psi}_a({\bf x})=\displaystyle\frac{1}{\sqrt{2}}
\left[ \eta_a({\bf x})+ \frac{\delta}{\delta \eta_a^*({\bf x})} \right], &
\nonumber \\
& \hat{\psi}_a^{\dagger}({\bf x})=\displaystyle\frac{1}{\sqrt{2}}
\left[ \eta_a^*({\bf x})+ \frac{\delta}{\delta \eta_a({\bf x})} \right], &
\end{eqnarray}
where $\eta_a({\bf x})$ and $\eta_a^*({\bf x})$ are anticommuting 
Grassmann numbers: 
\begin{equation}
\{ \eta_a({\bf x}),\eta_{a'}({\bf x}') \}=
\{ \eta_a^*({\bf x}),\eta_{a'}^*({\bf x}') \} =
\{ \eta_a({\bf x}),\eta_{a'}^*({\bf x}') \} =0. 
\end{equation}

Next
we introduce a complete orthonormal set of spinors $u_k({\bf x})$ and 
$v_k({\bf x})$. It is convenient to choose them to be equal 
to the spinors $u_k(x)$ and $v_k(x)$ at $t=0$, respectively. 
An arbitrary quantum state may be obtained by acting with the 
creation operators
\begin{eqnarray}\label{f10}
& \hat{b}_k^{\dagger}=\displaystyle\int d^3x \, 
\hat{\psi}^{\dagger}({\bf x}) u_k({\bf x}), & \nonumber \\
& \hat{d}_k^{\dagger}=\displaystyle\int d^3x \,
v_k^{\dagger}({\bf x}) \hat{\psi}({\bf x}),
\end{eqnarray}
on the vacuum $|0\rangle\equiv|\Psi_0\rangle$ represented by
\begin{equation}\label{f11}
\Psi_0[\eta,\eta^{\dagger}]=N\exp\left\{
\int d^3x \int d^3x'\, \eta^{\dagger}({\bf x}) \Omega({\bf x},{\bf x}')
\eta({\bf x}') \right\} .
\end{equation}
Here $\Omega({\bf x},{\bf x}')=\Omega^{(A)}({\bf x},{\bf x}')
-\Omega^{(P)}({\bf x},{\bf x}')$, $N$ is a constant chosen such that
$\langle \Psi_0|\Psi_0\rangle=1$ and the scalar product
is
\begin{equation}\label{scalprod}
\langle \Psi|\Psi'\rangle =
\int{\cal D}^2\eta\,
\Psi^*[\eta,\eta^{\dagger}]\Psi'[\eta,\eta^{\dagger}].
\end{equation}
In (\ref{scalprod}), 
${\cal D}^2\eta \equiv {\cal D}\eta{\cal D}\eta^{\dagger}$
and $\Psi^*$ is dual (not simply the complex conjugate
\cite{flor,kief}) to $\Psi$.  
The vacuum is chosen such that 
$\hat{b}_k \Psi_0=\hat{d}_k \Psi_0=0$. A functional 
$\Psi[\eta,\eta^{\dagger}]$ can be expanded as
\begin{equation}\label{f13}
\Psi[\eta,\eta^{\dagger}]=\sum_K c_K \Psi_K[\eta,\eta^{\dagger}],
\end{equation}
where the set $\{\Psi_K\}$ is a complete orthonormal set of 
Grassmann valued functionals. We choose this set 
such that each $\Psi_K$ is proportional 
to a functional of the form  
$\hat{b}_{k_1}^{\dagger}\cdots \hat{b}_{k_{n_P}}^{\dagger}
\hat{d}_{k'_1}^{\dagger}\cdots \hat{d}_{k'_{n_A}}^{\dagger} \Psi_0$, 
which 
means that each $\Psi_K$ has a definite number $n_P$ of particles 
and a definite number $n_A$ of antiparticles. Therefore, 
we can also write (\ref{f13}) as 
\begin{equation}\label{f14}
\Psi[\eta,\eta^{\dagger}]=\sum_{n_P,n_A=0}^{\infty} 
\tilde{\Psi}_{n_P,n_A}[\eta,\eta^{\dagger}].
\end{equation}
The tilde on $\tilde{\Psi}_{n_P,n_A}$ denotes that these 
functionals, in contrast with $\Psi$ and $\Psi_K$,
do not have unit norm.

Time-dependent states $\Psi[\eta,\eta^{\dagger},t]$ can be 
expanded as 
\begin{eqnarray}\label{f15}
\Psi[\eta,\eta^{\dagger},t] &=& \sum_K c_K(t) \Psi_K[\eta,\eta^{\dagger}]
\nonumber \\
&=& \sum_{n_P,n_A=0}^{\infty}
\tilde{\Psi}_{n_P,n_A}[\eta,\eta^{\dagger},t].
\end{eqnarray}
The time dependence of the c-number coefficients $c_K(t)$ is governed 
by the functional Schr\"odinger equation
\begin{equation}\label{f16}
H[\hat{\psi},\hat{\psi}^{\dagger}]\Psi[\eta,\eta^{\dagger},t]=
i\partial_t \Psi[\eta,\eta^{\dagger},t].
\end{equation}
Since the Hamiltonian $H$ is a hermitian operator, the norms 
\begin{equation}\label{f17}
\langle \Psi(t)|\Psi(t)\rangle = \sum_K |c_K(t)|^2
\end{equation}
of the states do not depend on time.
In particular, if $H$ is the free Hamiltonian
(i.e. the Hamiltonian that generates the second quantized free 
Dirac equation (\ref{f1})), then the quantities
$|c_K(t)|$ do not depend on time. This means that the 
average number 
of particles and antiparticles does not change with time  
when the interactions are absent.

Next we introduce the wave function of $n_P$ particles and
$n_A$ antiparticles, denoted as 
\begin{equation}\label{f18}  
\psi_{n_P,n_A}\equiv
\psi_{b_1\cdots b_{n_P}d_1\cdots d_{n_A}}
({\bf x}_1,\ldots,{\bf x}_{n_P},{\bf y}_1,\ldots,{\bf y}_{n_A},t).
\end{equation}
It has $n_P+n_A$ spinor indices $b_1,\cdots ,d_{n_A}$. 
For free fields, the 
(unnormalized) wave function can be calculated using the 
Heisenberg picture as
\begin{equation}\label{f19}
\psi_{n_P,n_A}=\langle 0| \hat{\psi}^{(P)}_{b_1}(t,{\bf x}_1)
\cdots \hat{\psi}^{(A)\dagger}_{d_{n_A}}(t,{\bf y}_{n_A})
|\Psi\rangle ,
\end{equation}
where $\hat{\psi}^{(P)}$ and $\hat{\psi}^{(A)}$ are extracted 
from $\hat{\psi}$ with the aid of (\ref{f5}). 
In the general, interacting case, the wave function can be calculated 
using the Schr\"odinger picture as
\begin{eqnarray}\label{f20}
\psi_{n_P,n_A} &=& \int{\cal D}^2\eta\, \Psi_0^*[\eta,\eta^{\dagger}]
e^{-i\varphi_0(t)}  \nonumber \\
& & \times
\hat{\psi}^{(P)}_{b_1}({\bf x}_1)
\cdots \hat{\psi}^{(A)\dagger}_{d_{n_A}}({\bf y}_{n_A})
\Psi[\eta,\eta^{\dagger},t].
\end{eqnarray}
Here the phase $\varphi_0(t)$ is defined by an expansion of the form 
of (\ref{f15}):
\begin{equation}\label{f21}
\hat{U}(t)\Psi_0[\eta,\eta^{\dagger}]=r_0(t)e^{i\varphi_0(t)}
\Psi_0[\eta,\eta^{\dagger}]
+\sum_{(n_P,n_A)\neq(0,0)}\ldots ,
\end{equation}
where $r_0(t)\!\geq\! 0$ and 
$\hat{U}(t)\! =\! U[\hat{\psi},\hat{\psi}^{\dagger}, t]$
is the unitary time-evolution operator that satisfies the 
Schr\"odinger equation (\ref{f16}). Note that only 
$\tilde{\Psi}_{n_P,n_A}$ from the expansion (\ref{f15}) 
contributes to (\ref{f20}).

The current attributed to the $i$-th corpuscle (particle or 
antiparticle) in the wave function $\psi_{n_P,n_A}$ is
\begin{equation}\label{f22}
j_{\mu(i)}=\left\{ 
\begin{array}{ll}
\bar{\chi}\gamma_{\mu(i)}\chi & \mbox{for particles,} \\ 
\bar{\chi}\gamma^T_{\mu(i)}\chi & \mbox{for antiparticles,}
\end{array}
\right.
\end{equation}
where we have used abbreviations
\begin{equation}\label{f23}
\bar{\chi}\Gamma_{(i)}\chi\equiv
\bar{\chi}_{a_1\cdots a_i\cdots a_n} (\Gamma)_{a_i a'_i}
\chi_{a_1\cdots a'_i\cdots a_n},
\end{equation}
\begin{eqnarray}\label{f24}
& \bar{\chi}_{b_1\cdots b_{n_P}d_1\cdots d_{n_A}}
\equiv \psi^*_{b'_1\cdots b'_{n_P}d_1\cdots d_{n_A}}
(\gamma_0)_{b'_1 b_1} \cdots (\gamma_0)_{b'_{n_P} b_{n_P}}, &
\nonumber \\
& \chi_{b_1\cdots b_{n_P}d_1\cdots d_{n_A}}        
\equiv (\gamma^T_0)_{d_1 d'_1} \cdots (\gamma^T_0)_{d_{n_A} d'_{n_A}}
\psi^*_{b_1\cdots b_{n_P}d'_1\cdots d'_{n_A}} , &
\end{eqnarray}
and the superscript $T$ denotes the transpose.
Eq.~(\ref{f22}) is the generalization of (\ref{f6}) for 
multiparticle wave functions. Therefore, the trajectory 
of the $i$-th corpuscle guided by the wave function 
$\psi_{n_P,n_A}$ is given by the generalization of 
(\ref{f7})
\begin{equation}\label{f25}
\frac{d{\bf x}_{(i)}}{dt}=\frac{{\bf j}_{(i)}}
{j_{0(i)}}.
\end{equation}

As an example, consider the wave function describing one positron
\begin{equation}\label{antiwf}
\psi^{(A)*}_a(x) = \psi_a(x) = 
\langle 0|\hat{\psi}^{(A)\dagger}_a(x)|\Psi\rangle .
\end{equation}
From (\ref{f24}) we find $\chi=\gamma_0^T\psi$, $\bar{\chi}=\psi^{\dagger}$, 
so (\ref{f22}) reduces to 
\begin{equation}\label{anticur}
j_{\mu}^{(A)}=\psi^{\dagger}\gamma^T_{\mu}\gamma_0^T\psi
=\psi^{(A)\dagger}\gamma_0\gamma_{\mu}\psi^{(A)}
=\bar{\psi}^{(A)}\gamma_{\mu}\psi^{(A)}.
\end{equation}
Note that the wave function that corresponds to the presence 
(not the absence) of a positron is $\psi_a^{(A)*}$ (not 
$\psi_a^{(A)}$). In this sense, we were slightly imprecise in
Sec.~\ref{S2} by referring to $\psi^{(A)}$ as the antiparticle 
part. Nevertheless, the expression (\ref{f6}) for the 
antiparticle current is consistent with (\ref{anticur}).   
The wave function (\ref{antiwf})
satisfies the same equation of motion (Dirac equation) as the 
field operator $\hat{\psi}^{(A)\dagger}_a(x)$. Consequently, the current 
(\ref{anticur}) is conserved 
\begin{equation}\label{conseq}
\partial^{\mu}j_{\mu}^{(A)}=0 ,
\end{equation}
and naturally interpreted as the positron current. In particular, 
the quantity $j_{0}^{(A)}=\psi^{(A)*}_a \psi^{(A)}_a$ is 
naturally interpreted as the positron density, which can be 
demonstrated as follows.
The operator of the positron density is \cite{nikol3}
\begin{equation}
\hat{n}^{(A)}(x)=\hat{\psi}^{(A)}_a(x) \hat{\psi}^{(A)\dagger}_a(x) .
\end{equation} 
Let $|\Psi\rangle$ be a free one positron state
\begin{equation}
|\Psi\rangle =\sum_k c_k \hat{d}_k^{\dagger} |0\rangle ,
\end{equation}
where $c_k$ are some coefficients. 
The corresponding wave function  
is given by the complex conjugate of
\begin{equation}
\psi^{(A)}(x)=\sum_k c_k^* v_k (x) ,
\end{equation}
while the expectation value of the positron density is 
\begin{equation}
\langle\Psi|\hat{n}^{(A)}(x)|\Psi\rangle =\psi^{(A)*}_a(x) \psi^{(A)}_a(x) .
\end{equation}

Note that the positron wave function is not a coefficient of a 
decomposition of a quantum-field state $|\Psi\rangle$
in eigenstates of the positron density operator. 
In the conventional 
interpretation of QFT, this would imply  
that the positron density cannot be interpreted as 
the positron {\it probability} density. Indeed, the conventional
interpretation of QFT does not have definite predictions on the 
probabilities of particle and antiparticle positions \cite{nikoldbb}. 
However, similarly to the bosonic case \cite{nikoldbb}, 
our interpretation of fermions is an {\em extension} 
of the conventional interpretation. The extension consists in 
postulating the particle and antiparticle trajectories. 
Eq.~(\ref{f25}), together with the 
conservation equation (\ref{conseq}), provides the consistency of the 
probabilistic interpretation, i.e., the equivariance. 
If the positrons in a statistical ensemble are initially distributed 
with probabilities equal to
$\psi^{(A)*}_a \psi^{(A)}_a$, then (\ref{f25}) and
(\ref{conseq}) imply that the positrons will be distributed according to 
$\psi^{(A)*}_a \psi^{(A)}_a$ at {\em all} times.   

It is also important to emphasize that, similarly to the bosonic case
\cite{nikoldbb}, 
our extension is compatible with the conventional interpretation 
in the sense that any definite prediction of the conventional 
interpretation is also a prediction of our interpretation.
With our extension, we do not deny any standard QFT effect 
that can be described purely in terms of quantum fields without explicit 
introduction of the concept of particles. 
In particular, although our extension 
does not attribute a particle trajectory to the vacuum, the standard 
measurable effects of the vacuum are not denied by our extension.
The set of all predictions of the conventional interpretation 
is a subset of the set of all predictions of our interpretation.

\section{BOSONIZATION OF FERMIONIC QFT}
\label{S4}

So far, we have not introduced any causal interpretation
of the fermionic fields $\hat{\psi}$ or $\eta$. Actually, 
these two quantities are not observables, so there is no 
need for a causal interpretation of them. However, 
we need a causal interpretation of the processes of 
creation and destruction of particles and antiparticles. 
For bosonic fields, this can be achieved by introducing 
a new causally evolving hidden variable - the effectivity 
of a particle \cite{nikoldbb} - 
determined by the causal evolution of 
bosonic fields. Unfortunately, one cannot introduce a 
similar causal theory for the Grassmann fields $\eta$ and 
$\eta^{\dagger}$, essentially because the quantity 
$\Psi^*[\eta,\eta^{\dagger},t]\Psi[\eta,\eta^{\dagger},t]$
is Grassmann valued, so, unlike to the 
bosonic case, one cannot interpret this quantity as a probability
density. Therefore, in this section we formulate another 
representation of fermionic states, more similar to that
of bosonic states. 
In Sec.~\ref{S6}, using the general method developed in 
Sec.~\ref{S5},
we use this representation to postulate 
a causal interpretation of fermionic QFT and a causal 
interpretation of the processes of 
creation and destruction of particles and antiparticles, 
similar to that of bosonic fields.   

Note first that the notion of the scalar product can be 
generalized in such a way that it may be Grassmann valued 
\cite{hall}, which allows us to write
$\Psi[\eta,\eta^{\dagger},t]=\langle\eta,\eta^{\dagger}|
\Psi(t)\rangle$ and
\begin{equation}\label{f26}
1=\int{\cal D}^2\eta \, 
|\eta,\eta^{\dagger}\rangle\langle\eta,\eta^{\dagger}| .
\end{equation}

Next, for each fermionic state $\Psi_K[\eta,\eta^{\dagger}]$ 
we introduce the corresponding bosonic state 
$\Psi_K[\varphi,\varphi^{\dagger}]$, with $\varphi({\bf x})$ and
$\varphi^{\dagger}({\bf x})$ being c-number 4-spinors.
The bosonic states $\Psi_K$ are obtained in the same way as 
the corresponding fermionic states, by acting with the 
bosonic creation operators $\hat{b}_k^{\dagger}$ 
and $\hat{d}_k^{\dagger}$ 
(that satisfy the corresponding commutation relations) 
on the bosonic vacuum
$\Psi_0[\varphi,\varphi^{\dagger}]$. (For details on the bosonic
functional methods, see e.g.~Refs.~\cite{holbook,flor,hall,long}).
Note that the set of bosonic states 
$\{\Psi_K[\varphi,\varphi^{\dagger}]\}$ does not contain 
states obtained by acting with the same creation operator 
on the vacuum more than once (e.g.
$(\hat{b}_k^{\dagger})^2|0\rangle$) because such states 
vanish identically in the fermionic case. Therefore, 
the set of states $\{\Psi_K[\varphi,\varphi^{\dagger}]\}$ 
is a complete orthonormal basis for a subspace of the 
whole space 
of all well-behaved functionals  
$\Phi[\varphi,\varphi^{\dagger}]$. 
On this subspace, the unit operator can be represented as
\begin{equation}\label{f27}
1= \sum_K |\Psi_K\rangle \langle\Psi_K| .
\end{equation}
We can also introduce the quantity
\begin{eqnarray}\label{f28}
\langle\varphi,\varphi^{\dagger}|\eta,\eta^{\dagger}\rangle
&=& 
\sum_K \langle\varphi,\varphi^{\dagger}|\Psi_K\rangle
\langle\Psi_K|\eta,\eta^{\dagger}\rangle
\nonumber \\
&=&
\sum_K \Psi_K[\varphi,\varphi^{\dagger}]
\Psi_K^*[\eta,\eta^{\dagger}],
\end{eqnarray}
so we see that the sets $\{\Psi_K[\eta,\eta^{\dagger}]\}$
and $\{\Psi_K[\varphi,\varphi^{\dagger}]\}$ are 
two representations of the same orthonormal basis 
$\{|\Psi_K\rangle\}$ for the same
Hilbert space of fermionic states.
In other words, the fermionic state $|\Psi(t)\rangle$ can be 
represented as $\Psi[\varphi,\varphi^{\dagger},t]=
\langle\varphi,\varphi^{\dagger}|\Psi(t)\rangle$, which, 
by using (\ref{f26}), (\ref{f28}) and (\ref{f15}), 
can be expanded as
\begin{eqnarray}\label{f29}
\Psi[\varphi,\varphi^{\dagger},t] &=& \sum_K c_K(t) 
\Psi_K[\varphi,\varphi^{\dagger}],
\nonumber \\
&=& \sum_{n_P,n_A=0}^{\infty}
\tilde{\Psi}_{n_P,n_A}[\varphi,\varphi^{\dagger},t].
\end{eqnarray} 
By inserting the unit operator 
$1=\int{\cal D}^2\varphi |\varphi,\varphi^{\dagger}\rangle  
\langle\varphi,\varphi^{\dagger}|$ in (\ref{f17}),
we see that the time-independent norm can be written as
\begin{equation}\label{f30}
\langle \Psi(t)|\Psi(t)\rangle =
\int{\cal D}^2\varphi\,
\Psi^*[\varphi,\varphi^{\dagger},t]\Psi[\varphi,\varphi^{\dagger},t]. 
\end{equation} 
Therefore, we interpret the quantity
\begin{equation}\label{f31}
\rho[\varphi,\varphi^{\dagger},t]=
\Psi^*[\varphi,\varphi^{\dagger},t]\Psi[\varphi,\varphi^{\dagger},t]
\end{equation}
as a positive definite probability density for spinors 
$\varphi$ and $\varphi^{\dagger}$ to have space dependence 
$\varphi({\bf x})$ and $\varphi^{\dagger}({\bf x})$, respectively,
at time $t$.

The Schr\"odinger equation (\ref{f16}) can also be written 
in the $\varphi$-representation as
\begin{equation}\label{f16phi}
\hat{H}_{\varphi}\Psi[\varphi,\varphi^{\dagger},t]=
i\partial_t \Psi[\varphi,\varphi^{\dagger},t].
\end{equation}
The Hamiltonian $\hat{H}_{\varphi}$ is defined by its 
action on wave functionals 
$\Psi[\varphi,\varphi^{\dagger},t]$. This action is determined by
\begin{eqnarray}\label{hamferbos}
\hat{H}_{\varphi}\Psi[\varphi,\varphi^{\dagger},t] & = &
\int {\cal D}^2\eta\int {\cal D}^2\varphi' 
\langle\varphi,\varphi^{\dagger}|\eta,\eta^{\dagger}\rangle 
\nonumber \\
& & \times
\hat{H} \langle\eta,\eta^{\dagger}|\varphi',\varphi'^{\dagger}\rangle
\Psi[\varphi',\varphi'^{\dagger},t],
\end{eqnarray}
where $\hat{H}=H[\hat{\psi},\hat{\psi}^{\dagger}]$ is the 
Hamiltonian in (\ref{f16}).

Having constructed the $\varphi$-representation of fermionic 
QFT, we can introduce a causal interpretation of it. 
We first develop
a general method of causal interpretation in the next section 
and then apply it to fermionic QFT in Sec.~\ref{S6}.

\section{GENERAL METHOD OF CAUSAL INTERPRETATION}
\label{S5}

In this section we develop a general method of causal 
interpretation of a quantum system described by a 
c-number valued wave function that satisfies a 
Schr\"odinger equation. For simplicity, 
we study the case of $n$ degrees of freedom denoted by 
a real $n$-dimensional vector $\vec{\varphi}$, but we write all 
equations in a form that can be easily generalized 
to the case with an infinite number of the degrees of freedom.

The wave function $\psi(\vec{\varphi},t)$ satisfies the 
Schr\"odinger equation
\begin{equation}\label{gen1}
\hat{H}\psi=i\partial_t \psi ,
\end{equation}
where $\hat{H}$ is an arbitrary hermitian Hamiltonian
(written in the $\vec{\varphi}$-representation). The quantity 
$\rho=\psi^*\psi$ is the probability density for the 
variables $\vec{\varphi}$. The 
corresponding average velocity is   
\begin{equation}\label{gen2}
\frac{d\langle\vec{\varphi}\rangle(t)}{dt}=
\int d^n\varphi \, \rho(\vec{\varphi},t)\vec{u}(\vec{\varphi},t) ,
\end{equation}
where $\vec{u}$ is defined as
\begin{equation}\label{gen3}
\vec{u}={\rm Re}\,i\frac{\psi^*[\hat{H},\vec{\varphi}]\psi}{\psi^*\psi}.
\end{equation}
We introduce the source $J$ defined as
\begin{equation}\label{gen4}
J=\frac{\partial\rho}{\partial t} +\vec{\nabla}(\rho\vec{u}).
\end{equation}
For many physically
interesting Hamiltonians, the source $J$ vanishes. However, 
in general, $J$ does not need to vanish. An example for which 
$J\neq 0$ is provided by the Hamiltonian 
$\hat{H}_{\varphi}$ defined by (\ref{hamferbos}).
We want to find the quantity $\vec{v}(\vec{\varphi},t)$ that 
has the property (\ref{gen2})
\begin{equation}\label{gen5}
\frac{d\langle\vec{\varphi}\rangle(t)}{dt}=
\int d^n\varphi \, \rho(\vec{\varphi},t)\vec{v}(\vec{\varphi},t) ,
\end{equation}   
but at the same time satisfies the equivariance property
\begin{equation}\label{gen6}
\frac{\partial\rho}{\partial t} +\vec{\nabla}(\rho\vec{v})=0.
\end{equation}
These two properties allow us to postulate 
a consistent causal interpretation
of quantum mechanics in which $\vec{\varphi}$ has definite values 
at each time $t$, determined by
\begin{equation}\label{gendbb}
\frac{d\vec{\varphi}}{dt}=\vec{v}(\vec{\varphi},t).
\end{equation}
In particular, the equivariance (\ref{gen6}) provides 
that the statistical distribution of the variables
$\vec{\varphi}$ is given by $\rho$ for any time $t$, provided 
that it is given by $\rho$ for some initial time $t_0$.
When $J=0$ then $\vec{v}=\vec{u}$, which corresponds to the 
usual dBB interpretation. The aim of this section is to 
generalize this to the general case $J\neq 0$.

Let us write the velocity $\vec{v}$ in the form
\begin{equation}\label{gen7}
\vec{v}=\vec{u}+\rho^{-1}\vec{\cal E},
\end{equation}
where $\vec{\cal E}(\vec{\varphi},t)$ is the quantity that needs to be 
determined. From (\ref{gen6}), we see that $\vec{\cal E}$ must be a 
solution of the equation
\begin{equation}\label{gen9}
\vec{\nabla}\vec{\cal E}=-J.
\end{equation}
Let $\vec{E}(\vec{\varphi},t)$ be some particular solution of 
(\ref{gen9}). Then
\begin{equation}\label{zv2}
\vec{\cal E}(\vec{\varphi},t)=\vec{e}(t)+\vec{E}(\vec{\varphi},t)
\end{equation}
is also a solution for an arbitrary $\vec{\varphi}$-independent 
function $\vec{e}(t)$. Comparing (\ref{gen5}) with (\ref{gen2}), we see that
we must require $\int\! d^n\varphi\,\vec{\cal E}=0$. This fixes the function 
$\vec{e}(t)$ to be
\begin{equation}\label{zv3}
\vec{e}(t)=-V^{-1}\int d^n\varphi\,\vec{E}(\vec{\varphi},t),
\end{equation}
where $V\equiv\int\! d^n\varphi$. 
It remains to choose the particular solution $\vec{E}$. We choose 
it such that $\vec{E}=0$ when $J=0$. Eqs.~(\ref{zv3}) and (\ref{zv2}) 
then imply that $\vec{\cal E}=0$ when $J=0$, i.e. that $\vec{v}=\vec{u}$ 
when $J=0$. There is still some arbitrariness in choosing the 
particular solution $\vec{E}$, so we proceed in a way that seems to be 
the simplest one. We take $\vec{E}$ to be a gradient of a scalar function
\begin{equation}\label{gen8}
\vec{E}=\vec{\nabla}\Phi,
\end{equation}
so $\Phi$ satisfies the Poisson equation
\begin{equation}\label{gen10}
\vec{\nabla}^2\Phi=-J.
\end{equation}
It is convenient to solve (\ref{gen10}) by using the Green-function 
method. By taking the solution to have the form
\begin{equation}\label{zv5}
\Phi(\vec{\varphi},t)=\int d^n \varphi'\, G(\vec{\varphi},\vec{\varphi}')
J(\vec{\varphi}',t),
\end{equation}
we see from (\ref{gen10}) that $G$ satisfies
\begin{equation}\label{zv6}
\vec{\nabla}^2 G(\vec{\varphi},\vec{\varphi}')=
-\delta^n(\vec{\varphi}-\vec{\varphi}').
\end{equation}
The solution of (\ref{zv6}) is most easily expressed as a 
Fourier transform
\begin{equation}
G(\vec{\varphi},\vec{\varphi}')=\displaystyle\int \frac{d^n k}{(2\pi)^n}
\frac{e^{i\vec{k}(\vec{\varphi}-\vec{\varphi}')}}{\vec{k}^2}.
\end{equation}
To eliminate the factor $1/(2\pi)^n$, we 
introduce a new integration variable $\vec{\chi}=\vec{k}/2\pi$. 
In this way, (\ref{zv5}) becomes
\begin{equation}\label{gen13}
\Phi(\vec{\varphi},t)=\displaystyle\int d^n \chi
\displaystyle\int d^n \varphi' \, 
\frac{e^{i2\pi\vec{\chi}(\vec{\varphi}-\vec{\varphi}')}}
{(2\pi)^2 \vec{\chi}^2} J(\vec{\varphi}',t) ,
\end{equation}
which is 
written in the form appropriate for a generalization to an infinite 
number of the degrees of freedom.

The formalism above can also be easily generalized to the case 
in which the integration measure in (\ref{gen2}) is not simply 
$d^n\varphi$, but a more general expression of the form 
$d^n\varphi\, \mu(\vec{\varphi},t)$. 
In some cases, $\mu$ is a Jacobian that 
can be eliminated by a redefinition of the integration variables 
$\vec{\varphi}$. If this is not the case, $\mu$ can always be absorbed 
into a redefinition of the density $\rho\rightarrow\mu\rho$.

There is one additional tacit assumption in our general analysis 
that does not need to be satisfied in an even more general
situation: the components $\varphi_i$ of the vector 
$\vec{\varphi}$ are assumed to be continuous parameters. 
If these parameters were
discrete instead of continuous, then Eq.~(\ref{gen2}) would 
contain a summation over allowed values of $\vec{\varphi}$
instead of the integration over them. Examples of discrete 
variables are the $z$-component of spin and the   
momentum of a particle confined in a finite volume.  
There exists a consistent causal interpretation for these two 
examples \cite{holbook}, but it is not clear how to 
introduce a general systematic method of 
causal interpretation of discrete quantum variables.
Formally, one might reduce the summation to an integration by 
introducing the integration measure in which the function $\mu$ 
is essentially a sum of $\delta$-functions. However, owing to the 
singular nature of $\delta$-functions, it is not clear 
that this would lead to a mathematically well-defined causal 
theory (see e.g.~Ref.~\cite{nikoldelta}). A recent approach \cite{hym} 
suggests a solution of this problem based on smearing of the 
$\delta$-functions, but this approach does not seem to be unique.
Therefore, the formulation of a general method of causal 
interpretation of discrete quantum variables is still 
a challenge. Fortunately, we do not need it in this paper.

\section{CAUSAL INTERPRETATION OF FERMIONIC QFT}
\label{S6}

We are now ready to formulate a causal interpretation of fermionic QFT.
In Sec.~\ref{S4} we have been working with complex spinors $\varphi$ and 
$\varphi^{\dagger}$, having complex components $\varphi_a$ and 
$\varphi_a^*$, respectively. On the other hand, in Sec.~\ref{S5} we 
have developed a general method of causal interpretation for real 
variables. Therefore, in this section
we first introduce real spinor  
components $\varphi_a^1$ and $\varphi_a^2$ defined by 
$\varphi_a=(\varphi_a^1+i\varphi_a^2)/\sqrt{2}$, so that
\begin{equation}                                   
\varphi_a^1=\displaystyle\frac{\varphi_a^* +\varphi_a}{\sqrt{2}} ,     
\;\;\;\;\;
\varphi_a^2=i\displaystyle\frac{\varphi_a^* -\varphi_a}{\sqrt{2}} . 
\end{equation}
Each functional of the complex spinors $\varphi$ and $\varphi^{\dagger}$
can also be viewed as a functional of the real spinors  $\varphi^1$ and
$\varphi^2$. 
To simplify the notation, when appropriate, 
we denote the functionals of the 
form of $A[\varphi^1,\varphi^2,t,{\bf x}]$ by a shorter 
simbol $A[{\bf x}]$.
 
Following the general method developed 
in Sec.~\ref{S5}, we introduce the quantity
\begin{equation}
u_a^r[{\bf x}]={\rm Re}\,i
\displaystyle\frac{\Psi^*[\hat{H}_{\varphi},\varphi_a^r({\bf x})]\Psi}
{\Psi^*\Psi} , 
\end{equation}
where $r=1,2$ and $\Psi=\Psi[\varphi^1,\varphi^2,t]$. 
Next, we introduce the source 
\begin{equation}
J=\frac{\partial\rho}{\partial t}+\sum_{a,r}\int d^3x 
\frac{\delta(\rho u_a^r[{\bf x}])}{\delta \varphi_a^r({\bf x})} ,
\end{equation}
where $\rho=\Psi^*\Psi$. Introducing the notation
\begin{equation}
\alpha\cdot\beta\equiv\sum_{a,r}\int d^3x \,
\alpha_a^r({\bf x})\beta_a^r({\bf x}),
\end{equation}
Eq.~(\ref{gen13}) generalizes to 
\begin{equation}
\Phi[\varphi^1,\varphi^2,t]=
\displaystyle\int {\cal D}^2\chi
\displaystyle\int {\cal D}^2 \varphi' \,
\frac{e^{i2\pi\chi\cdot(\varphi-\varphi')}}
{(2\pi)^2 \chi\cdot\chi} \, J[\varphi'^1,\varphi'^2,t] .
\end{equation}
Introducing the quantities
\begin{equation}
E_a^r[{\bf x}]=
\frac{\delta\Phi}{\delta \varphi_a^r({\bf x})} , 
\end{equation}
\begin{equation}
e_a^r(t,{\bf x})=-V^{-1}\displaystyle\int {\cal D}^2 \varphi\, 
E_a^r[\varphi^1,\varphi^2,t,{\bf x}], 
\end{equation}
where $V\equiv\int {\cal D}^2 \varphi$,
the corresponding velocity is given by
\begin{equation}
v_a^r[{\bf x}]=u_a^r[{\bf x}]+\rho^{-1}(e_a^r(t,{\bf x})
+E_a^r[{\bf x}]).
\end{equation}
Now we introduce hidden variables 
$\varphi^1(t,{\bf x})$ and $\varphi^2(t,{\bf x})$. 
The causal evolution of these 
hidden variables is given by
\begin{equation}\label{f35}
\displaystyle\frac{\partial\varphi_a^r(t,{\bf x})}{\partial t}=
v_a^r[\varphi^1,\varphi^2,t,{\bf x}], 
\end{equation}
where it is understood that the right-hand side is calculated 
at $\varphi^r({\bf x})=\varphi^r(t,{\bf x})$. 

In analogy with ordinary bosonic fields \cite{nikoldbb}, 
we also introduce the effectivity $e_{n_P,n_A}$
of the particles guided 
by the wave function $\psi_{n_P,n_A}$, given by
\begin{equation}\label{f36}
e_{n_P,n_A}[\varphi^1,\varphi^2,t]=
\frac{|\tilde{\Psi}_{n_P,n_A}[\varphi^1,\varphi^2,t]|^2}
{\displaystyle\sum_{n'_P,n'_A}
|\tilde{\Psi}_{n'_P,n'_A}[\varphi^1,\varphi^2,t]|^2}.
\end{equation}
Since the physical meaning of the effectivity is discussed 
in detail in Ref.~\cite{nikoldbb} for ordinary bosonic 
fields, here we only note the final results valid also for 
the bosonic representation of the fermionic states above. 
In an ideal experiment in which the number of particles is measured, 
different $\tilde{\Psi}_{n_P,n_A}$ functionals do not overlap in the 
$(\varphi^1,\varphi^2)$ space. Consequently, the fields 
$\varphi^1$ and $\varphi^2$ necessarily enter one and 
only one of the ``channels" $\tilde{\Psi}_{n_P,n_A}$. In this case, 
the effectivity (\ref{f36}) of the corresponding particles 
is equal to 1, while that of all other particles guided by 
other wave functions $\psi_{n'_P,n'_A}$ is equal to 0. 
The effect is the same as if the wave functional $\Psi$ ``collapsed" into
one of the states $\tilde{\Psi}_{n_P,n_A}$ with a definite 
number of particles and antiparticles.
This provides a causal description of the creation of particles
guided by $\psi_{n_P,n_A}$ and the destruction of all other particles 
guided by other wave functions $\psi_{n'_P,n'_A}$.

Note that {\em both} the corpuscle (particle and antiparticle) positions and 
the commuting fields 
$\varphi^1$, $\varphi^2$ are the preferred hidden variables that describe 
the elements of reality associated with fermions. This would not be possible 
if the corpuscle positions were described by hermitian operators that do not 
commute with $\varphi^1$, $\varphi^2$.
However,
just as in the relativistic bosonic case \cite{nikoldbb}, 
there are {\em no} hermitian corpuscle-position {\em operators} 
associated with the corpuscle positions. 
At the level of first-quantization formalism, the absence of 
the hermitian corpuscle-position 
operators is a consequence of the fact that the 
separate sets $\{ u_k ({\bf x}) \}$ and $\{ v_k ({\bf x}) \}$ are not
complete, so one cannot construct a $\delta$-function as a linear 
combination of the functions of only one set. Similarly, at the 
level of second-quantization formalism, the are no 
1-particle states $|\Psi^{(P)}\rangle$ 
with the property 
$\hat{n}^{(P)}({\bf x})|\Psi^{(P)}\rangle =\delta^3({\bf
x})|\Psi^{(P)}\rangle$ (and analogously for 1-antiparticle states).
These facts are also related to the 
already mentioned fact that the conventional theory, in general, 
does not give definite predictions on the probabilities of 
corpuscle positions. 
In our theory, the corpuscle positions correspond to an {\em extension} 
of the conventional theory. As discussed 
in Sec.~\ref{S3}, the consistency of the probabilistic interpretation 
of particle and antiparticle wave functions rests on the fact that 
the probabilities correspond only to an effective 
statistical description of the underlying
deterministic Bohmian trajectories. 
 

\section{CONCLUSION}
\label{S7}

In this paper, three new results 
related to the causal interpretation of fermions 
have been obtained. First, the usual causal interpretation 
of the Dirac equation is modified such that it makes a difference 
between particle and antiparticle trajectories. Second, fermionic 
interacting
QFT is used to construct the multiparticle wave functions that describe 
particle and antiparticle trajectories in systems with many particles.
Third, a causal interpretation of fermionic fields themselves, 
based on bosonization of fermionic fields, is 
constructed and used to formulate a causal interpretation of 
particle creation and destruction. 

Besides this progress in the causal interpretation of fermions, 
a general method of causal interpretation is developed. 
It can be applied to any quantum system described by a c-number valued 
wave function $\psi(\vec{\varphi},t)$
satisfying a Schr\"odinger equation, provided that the 
c-number parameters $\varphi_i$ are continuous.
 
\vspace{0.4cm}
\noindent
{\bf Acknowledgements.}
The author is grateful to R.~Tumulka for valuable comments 
and discussions and to an anonymous referee for valuable 
critical remarks that improved the paper. 
This work was supported by the Ministry of Science and Technology of the
Republic of Croatia under Contract No.~0098002.

\end{document}